\documentclass{article}
\usepackage{amsmath}
\usepackage{cite}
\usepackage{graphicx}
\usepackage{dcolumn}

\begin{document}

\date{}
\title{On the two-dimensional harmonic oscillator with an electric field confined
to a circular box}
\author{Francisco M. Fern\'{a}ndez\thanks{%
fernande@quimica.unlp.edu.ar} \\
INIFTA, DQT, Sucursal 4, C. C. 16, \\
1900 La Plata, Argentina \\
Javier Garcia \\
Instituto de F\'{i}sica La Plata,\\
Consejo Nacional de Investigaciones Cient\'{i}ficas y T\'{e}cnicas, and\\
Departamento de F\'{i}sica, Universidad Nacional de La Plata,\\
C.C. 67, 1900, La Plata, Argentina \\
Norberto Aquino \\
Departamento de F\'isica, \\
Universidad Aut\'onoma Metropolitana-Iztapalapa, \\
Av. Ferrocarril de San Rafael Atlixco 186, \\
Iztapalapa, C.P. 09310 Cd. Mexico, Mexico\\
Antonio Flores-Riveros\\
Instituto de F\'{i}sica, \\
Benem\'{e}rita Universidad Aut\'{o}noma de Puebla,\\
Av. San Claudio y Blvd. 18 Sur, Col. San Manuel, \\
Puebla, Pue, M\'{e}xico, C. P. 72570}
\maketitle

\begin{abstract}
We revisit the quantum-mechanical two-dimensional harmonic oscillator with
an electric field confined to a circular box of impenetrable walls. In order
to obtain the energy spectrum we resort to the Rayleigh-Ritz method with
polynomial and Gaussian basis sets. We compare present results with those
derived recently by other authors. We discuss the limits of large and small
box radius and also do some calculations with perturbation theory.
\end{abstract}

\section{Introduction}

\label{sec:intro}

In a paper published recently Cruz et al\cite{CAPF24} studied a model for an
electron confined to a circle of radius $r_{0}$ under the effect of a
harmonic interaction and a uniform electric field. They solved the
Schr\"{o}dinger equation by means of the linear variational method (commonly
known as the Rayleigh-Rith method (RRM)\cite{P68,SO96}) and studied the
effect of the box size and the magnitude of the electric field on the
Shannon entropy and Fisher information for some of the lowest states of the
system.

In their calculation Cruz et al resorted to an incomplete trial function
that is a linear combination of states sharing the same radial part. Since
there are no other results that may be chosen as benchmark it was not
possible to estimate the accuracy of those variational results. In this
paper we repeat the calculation using a suitable basis set. Since the RRM
yields increasingly accurate upper bounds to the all eigenvalues of the
problem\cite{M33,F22} we can easily estimate the accuracy of the results of
Cruz et al.

\section{Two-dimensional harmonic oscillator confined to a circle with a
constant electric field}

\label{sec:CHO2DEF}

In what follows we outline some general features of the Schr\"{o}dinger
equation for the model as well as the earlier calculation carried out by
Cruz et al\cite{CAPF24}.

\subsection{Some general features of the eigenvalue equation}

\label{subsec:general_features}

The problem discussed by Cruz et al is given by the Hamiltonian operator
\begin{equation}
H=-\frac{\hbar ^{2}}{2m}\nabla ^{2}+\frac{k}{2}r^{2}+efx+V_{c}(r),
\label{eq:H_initial}
\end{equation}
where $m$ is the electron mass, $e$ is the electron charge, $k$ is the
oscillator force constant, $f$ is the magnitude of the electric field and $%
V_{c}(r)$ is a confining potential given by
\begin{equation}
V_{c}(r)=\left\{
\begin{array}{l}
0\mathrm{\ if\;}0\leq r<r_{0} \\
\infty \;\mathrm{if\;}r\geq r_{0}
\end{array}
\right. ,  \label{eq:V_c}
\end{equation}
that forces the boundary condition $\psi (x,y)=0$ for all $r=\sqrt{%
x^{2}+y^{2}}\geq r_{0}$.

\subsection{Dimensionless equations}

\label{subsec:dimensionless}

Before discussing the solution of the Schr\"{o}dinger equation with the
Hamiltonian operator (\ref{eq:H_initial}) we briefly outline some relevant
features of the model. To begin with, we consider the transformation $%
(x,y)=(L\tilde{x},L\tilde{y})$, $\nabla ^{2}=L^{-2}\tilde{\nabla}^{2}$,
where $L$ is an arbitrary unit of length and $(\tilde{x},\tilde{y})$ are
dimensionless cartesian coordinates. In this way we derive the dimensionless
Hamiltonian operator\cite{F20} (we omit the confining potential from now on)
\begin{equation}
\tilde{H}=\frac{mL^{2}}{\hbar ^{2}}H=-\frac{1}{2}\tilde{\nabla}^{2}+\frac{%
mkL^{4}}{2\hbar ^{2}}\tilde{r}^{2}+\frac{mefL^{3}}{\hbar ^{2}}\tilde{x}.
\label{eq:H_transf}
\end{equation}
The boundary condition now becomes $\tilde{\psi}(\tilde{x},\tilde{y})=0$ for
all $\tilde{r}\geq \tilde{r}_{0}$, $\tilde{r}=r/L$, $\tilde{r}_{0}=r_{0}/L$.

We first consider the case $L=r_{0}$ that leads to
\begin{equation}
\tilde{H}=\frac{mr_{0}^{2}}{\hbar ^{2}}H=-\frac{1}{2}\tilde{\nabla}^{2}+%
\frac{mkr_{0}^{4}}{2\hbar ^{2}}\tilde{r}^{2}+\frac{mefr_{0}^{3}}{\hbar ^{2}}%
\tilde{x},  \label{eq:H_L=r0}
\end{equation}
from which it follows that
\begin{equation}
\lim\limits_{r_{0}\rightarrow 0}\tilde{H}=-\frac{1}{2}\tilde{\nabla}^{2}.
\label{eq:H_dim_r0->0}
\end{equation}
That is to say, when $r_{0}\rightarrow 0$ we obtain the model of a particle
in a circular box with radius $\tilde{r}_{0}=1$. We will come back to this
issue later on.

Cruz et al chose $L=\sqrt{\hbar /\sqrt{mk}}=\sqrt{\hbar /m\omega }$, where $%
\omega =\sqrt{k/m}$, so that
\begin{equation}
\tilde{H}=\frac{H}{\hbar \omega }=-\frac{1}{2}\tilde{\nabla}^{2}+\frac{1}{2}%
\tilde{r}^{2}+\lambda \tilde{x},\;\lambda =\frac{mefL^{3}}{\hbar ^{2}}.
\label{eq:H_LHO}
\end{equation}
Throughout this paper we will use this form of the dimensionless Hamiltonian
operator and omit the tilde on the dimensionless variables.

\subsection{Earlier calculations}

\label{subsec:earlier_calc}

According to the results of the preceding subsection the dimensionless
Hamiltonian operator for this model can be written as
\begin{equation}
H=-\frac{1}{2}\nabla ^{2}+\frac{1}{2}r^{2}+\lambda r\cos (\phi ),
\label{eq:H_CHO2DEF}
\end{equation}
where $r=\sqrt{x^{2}+y^{2}}$ and $0\leq \phi <2\pi $. The eigenfunctions $%
\psi (r,\phi )$ satisfy the boundary condition $\psi (r_{0},\phi )=0$.

Cruz et al first solved the eigenvalue equation for $\lambda =0$ and
obtained the eigenfunctions in polar coordinates $\psi _{nm}^{(0)}\left(
r,\phi \right) =R_{nm}(r)\frac{e^{im\phi }}{\sqrt{2\pi }}$, where $%
n=0,1,\ldots $ and $m=0,\pm 1,\pm 2,\ldots $ are the radial and magnetic
quantum numbers, respectively. With such solutions they proposed trial
variational functions of the form
\begin{equation}
\Psi (r,\phi )=\sum_{m=-M}^{M}c_{m}\psi _{nm}^{(0)}(r,\phi ),
\label{eq:trial_Psi_inc}
\end{equation}
where $\psi _{nm}^{(0)}$ are the eigenfunctions of the $2D$ confined
harmonic oscillator\cite{CAPF24,MCA10}. It is clear that the general trial
function should be
\begin{equation}
\Psi (r,\phi )=\sum_{n=0}^{N}\sum_{m=-M}^{M}c_{nm}\psi _{nm}^{(0)}(r,\phi ),
\label{eq:trial_Psi_comp}
\end{equation}
because the term $\lambda r\cos (\phi )$ clearly couples states with
different values of $n$. The latter approach requires the calculation of
integrals of the form
\begin{equation}
\int_{0}^{r_{0}}R_{nm}(r)rR_{n^{\prime }m^{\prime }}(r)r\,dr,
\end{equation}
that we can avoid if we choose an alternative basis set.

Before discussing an alternative implementation of the RRM we first consider
some relevant features of the solutions of the Schr\"{o}dinger equation with
the Hamiltonian operator (\ref{eq:H_CHO2DEF}). Since the Hamiltonian is
invariant under the transformation $\phi \rightarrow -\phi $, then the
eigenfunctions are either even ($\psi ^{(e)}(r,-\phi )=\psi ^{(e)}(r,\phi )$%
) or odd ($\psi ^{(o)}(r,-\phi )=-\psi ^{(o)}(r,\phi )$). For this reason,
the solutions at $\lambda =0$ can be more conveniently written as
\begin{eqnarray}
\psi _{n\nu }^{(0,e)}\left( r,\phi \right) &=&R_{n\nu }^{(e)}(r)\cos (\nu
\phi ),\;\nu =|m|=0,1,\ldots ,  \nonumber \\
\psi _{n\nu }^{(0,o)}\left( r,\phi \right) &=&R_{n\nu }^{(o)}(r)\sin (\nu
\phi ),\;\nu =|m|=1,2,\ldots ,  \label{eq:psi_lamb=0_e,o}
\end{eqnarray}
and we can treat them separately by means of the RRM. Note that there is no
degeneracy within each set of solutions and that Cruz et al arrived at the
same two sets of basis functions empirically.

The second relevant feature is given by the unitary transformation $U\phi
U^{\dagger }=\phi +\pi $ that leads to $UH(\lambda )U^{\dagger }=H(-\lambda
) $; therefore,
\begin{equation}
UH(\lambda )\psi =UH(\lambda )U^{\dagger }U\psi =H(-\lambda )U\psi
=E(\lambda )U\psi .
\end{equation}
Since we have removed any degeneracy within the sets of even and odd
solutions we conclude that $E(\lambda )=E(-\lambda )$; that is to say, the
eigenvalues are even functions of $\lambda $.

\section{Rayleigh-Ritz method}

\label{sec:RR}

Since the RRM is well known\cite{P68,SO96,M33,F24}, we do not discuss it
here in detail. In what follows we simply describe some basis sets that, in
our opinion, are suitable for a successful calculation. The procedure
described in this section is motivated by a recent discussion of the RRM
with a non-orthogonal basis set\cite{F24}.

\subsection{Polynomial basis set}

\label{subsec:polybas}

The simplest basis set is undoubtedly a polynomial one. For the even and odd
functions we propose the non-orthogonal basis sets
\begin{eqnarray}
f_{i,j}^{(e)}(r,\phi ) &=&r^{i}(r_{0}-r)\cos (j\phi ),\;i=0,1,\ldots
,\;j=0,1,\ldots ,i,  \nonumber \\
f_{i,j}^{(o)}(r,\phi ) &=&r^{i}(r_{0}-r)\sin (j\phi ),\;i=1,2,\ldots
,\;j=1,2,\ldots ,i.  \label{eq:f^eo}
\end{eqnarray}
The calculation of the matrix elements $S_{iji^{\prime }j^{\prime
}}=\left\langle f_{ij}\right| \left. f_{i^{\prime }j^{\prime }}\right\rangle
$ and $H_{iji^{\prime }j^{\prime }}=\left\langle f_{ij}\right| H\left|
f_{i^{\prime }j^{\prime }}\right\rangle $ is straightforward and we can
easily apply the RRM\cite{P68,SO96,M33,F24} after converting $S_{iji^{\prime
}j^{\prime }}$ and $H_{iji^{\prime }j^{\prime }}$ into two-dimensional
arrays. The main disadvantage of these basis sets is that they are not
expected to be suitable for large values of $r_{0}$ because they do not
exhibit the correct Gaussian behaviour when $r_{0}\rightarrow \infty $.
However, they are suitable for most of our calculations.

We choose $j=0,1,\ldots ,N$, $i=j,j+1,\ldots ,N$ for even states and $%
j=1,2,\ldots ,N$, $i=j,j+1,\ldots ,N$ for odd ones.\ Consequently, the
number of basis functions is $(N+1)(N+2)/2$ in the former case and $N(N+1)/2$
in the latter.

Tables \ref{tab:e1_0.05} and \ref{tab:o1_0.05} show the convergence of the
RRM for the first four even and first four odd eigenvalues, respectively,
when $r_{0}=1$ and $\lambda =0.05$. Both tables also show the eigenvalues
obtained by Cruz et al. It is clear that the results of the latter authors
are reasonably accurate at least for small values of $r_{0}$ and $\lambda $.
Table~\ref{tab:Er0lam} shows present results for all the values of $r_{0}$
and $\lambda $ considered by Cruz et al. The discrepancy between present
eigenvalues and those in table 1 of Cruz et al\cite{CAPF24} is noticeable
for large values of $r_{0}$ and $\lambda $. The reason is that the ansatz
used by those authors, shown above in equation (\ref{eq:trial_Psi_inc}),
requires the contribution of basis functions $\psi _{n^{\prime }m}^{(0)}$
with $n^{\prime }\neq n$ as shown in equation (\ref{eq:trial_Psi_comp}).

The analysis carried out in subsection~\ref{subsec:dimensionless} predicts
that

\begin{equation}
\lim\limits_{r_{0}\rightarrow 0}r_{0}^{2}E_{n\nu }=E_{n\nu }^{PB},
\label{eq:r0^2E_r0_zero}
\end{equation}
where $E_{n\nu }^{PB}$ is the corresponding eigenvalue of the particle in a
box of radius $r_{0}=1$. Table~\ref{tab:r0^2E} shows that $r_{0}^{2}E_{n\nu
}\approx E_{n\nu }^{PB}$ for $r_{0}=0.01$.

\subsection{Gaussian basis set}

\label{subsec:Gaussbas}

When $r_{0}\rightarrow \infty $ the problem is separable in cartesian
coordinates, the Hamiltonian can be written as
\begin{equation}
H=-\frac{1}{2}\nabla ^{2}+\frac{1}{2}\left( x+\lambda \right) ^{2}+\frac{1}{2%
}y^{2}-\frac{\lambda ^{2}}{2}  \label{eq:H_r0->inf}
\end{equation}
and we conclude that
\begin{equation}
\lim\limits_{r_{0}\rightarrow \infty }E_{n\nu }\left( r_{0},\lambda \right)
=n_{1}+n_{2}+1-\frac{\lambda ^{2}}{2},\;n_{1},n_{2}=0,1,\ldots
\label{eq:E_r0->inf}
\end{equation}
Note that we have decided to label the eigenvalues with the quantum numbers $%
n$ and $\nu $ that are natural for the unperturbed problem with $\lambda =0$%
. Upon solving the Schr\"{o}dinger equation for $r_{0}\rightarrow \infty $
and $\lambda =0$ we realize that $n_{1}+n_{2}=2n+\nu $. The energy levels (%
\ref{eq:E_r0->inf}) depend on $n_{1}+n_{2}$ and are $\left(
n_{1}+n_{2}+1\right) $-fold degenerate.

As $r_{0}$ increases, the rate of convergence of the RRM with the basis set (%
\ref{eq:f^eo}) becomes slower. In order to improve the performance of the
approach for large values of $r_{0}$ we can choose
\begin{eqnarray}
g_{i,j}^{(e)}(r,\phi ) &=&r^{i}(r_{0}-r)e^{-r^{2}/2}\cos (j\phi
),\;i=0,1,\ldots ,\;j=0,1,\ldots ,i,  \nonumber \\
g_{i,j}^{(o)}(r,\phi ) &=&r^{i}(r_{0}-r)e^{-r^{2}/2}\sin (j\phi
),\;i=1,2,\ldots ,\;j=1,2,\ldots ,i.  \label{eq:g^(eo)}
\end{eqnarray}
The effect of a suitable asymptotic behaviour of the basis functions on the
rate of convergence of the RRM has been discussed earlier for some
anharmonic oscillators\cite{FG14}.

Table~\ref{tab:E00(r0)} shows $E_{00}\left( r_{0},\lambda \right) $ for
increasing values of $r_{0}$ and some values of $\lambda $. It is clear that
the difference $E_{00}\left( r_{0},\lambda \right) -E_{00}\left( \infty
,\lambda \right) $ for a given value of $r_{0}$ increases with $\lambda $.

\section{Perturbation theory}

\label{sec:PT}

In this section we apply perturbation theory to the calculation of the
energies of the model. We will discuss two alternative implementations of
the approximate method.

\subsection{Standard perturbation theory}

\label{subsec:standard_PT}

In order to apply perturbation theory to the present model we define $H_{0}=-%
\frac{1}{2}\nabla ^{2}+\frac{1}{2}r^{2}$ and $H^{\prime }=r\cos (\phi )$.
Since the eigenvalues are even functions of $\lambda $ the perturbation
series should be
\begin{equation}
E_{n\nu }=\sum_{j=0}^{\infty }E_{n\nu }^{(2j)}\lambda ^{2j}.
\label{eq:PT_series}
\end{equation}
The first correction is given by
\begin{equation}
E_{n\nu }^{(2)}=\sum_{n^{\prime }=0}^{\infty }\sum_{|\nu ^{\prime }-\nu |=1}%
\frac{\left| \left\langle \psi _{n\nu }^{(0)}\right| r\cos (\phi )\left|
\psi _{n^{\prime }\nu ^{\prime }}^{(0)}\right\rangle \right| ^{2}}{E_{n\nu
}^{(0)}-E_{n^{\prime }\nu ^{\prime }}^{(0)}}  \label{eq:E^(2)}
\end{equation}
where $\psi _{n\nu }^{(0)}$ and $E_{n\nu }^{(0)}$ are the
eigenfunctions and eigenvalues of $H_{0}$\cite{CAPF24,MCA10}. It
should be taken into account that the unperturbed eigenfunctions
and eigenvalues in this calculation are those indicated in
equation (\ref{eq:psi_lamb=0_e,o}) and that the even and odd
states are treated separately. Since the terms in this sum
decrease quite rapidly just a few of them are enough for a
reasonable accuracy. For example:
\begin{equation}
E_{00}^{(2)}=-0.023766061092-0.0000279159-1.0265\times 10^{-6}=-0.0237950030,
\label{eq:E_(00)^(2)}
\end{equation}
for $r_{0}=1$. The approximate ground-state energy for this box radius is
\begin{equation}
E_{00}=3-0.0237950030\lambda ^{2}+\mathcal{O}\left( \lambda ^{4}\right) .
\label{eq:E_(00)_PT}
\end{equation}
In the next subsection we will derive more accurate perturbation results by
means of an alternative approach and discuss the accuracy of second-order
perturbation theory.

\subsection{Alternative perturbation approach}

\label{subsec:alternative_PT}

Instead of the eigenfunctions and eigenvalues of $H_{0}$ derived in the way
indicated by Cruz et al\cite{CAPF24} and Montgomery et al\cite{MCA10} we
obtained them by means of the RRM with the Gaussian basis set (\ref
{eq:g^(eo)}) and obtained the perturbation corrections systematically
through the equations given in chapter 1 of one of the available books on
perturbation theory\cite{F01}. In this way we obtained many perturbation
coefficients for several states. In table~\ref{tab:PT_E_n} we show $E^{(0)}$
and $E^{(2)}$ for the first five states when $r_{0}=1$. Note that the
splitting between the fourth and fifth states occurs at higher perturbation
orders.

\begin{figure}[tbp]
\begin{center}
\includegraphics[width=9cm]{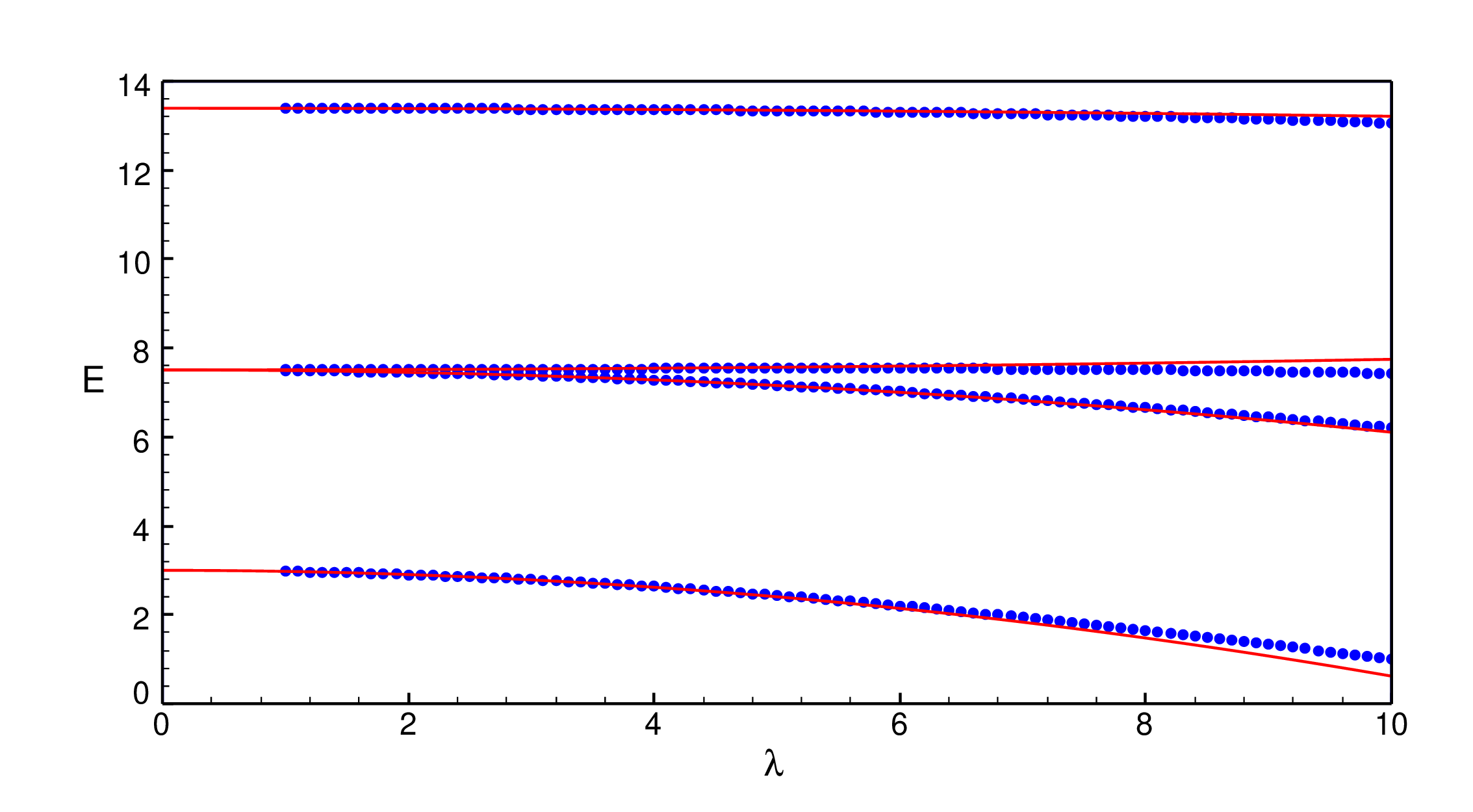}
\end{center}
\caption{Eigenvalues for the first five states calculated by means of the
RRM (blue points) and perturbation theory (red line)}
\label{Fig:En_PT_r0_1}
\end{figure}

Figure~\ref{Fig:En_PT_r0_1} shows the first five eigenvalues calculated by
second-order perturbation theory and by means of the RRM. The agreement is
satisfactory in a wide range of values of $\lambda $.

From the results of subsection~\ref{subsec:Gaussbas} we conclude that
\begin{equation}
\lim\limits_{r_{0}\rightarrow \infty }E_{n\nu }^{(2)}=-\frac{1}{2}.
\label{eq:E^(2)_r0_inf}
\end{equation}
Table~\ref{tab:E^(2)_(nn)} shows that $E_{n\nu }^{(0)}$ tends to $2n+\nu +1$
and $E_{n\nu }^{(2)}$ to $-0.5$ as $r_{0}$ increases.

\section{Conclusions}

\label{sec:conclusions}

We have shown that the incomplete basis set used by Cruz et al\cite{CAPF24}
is not suitable for large values of $r_{0}$ and $\lambda $. We proposed two
alternative basis sets of functions; one of them leads to simpler matrix
elements but is rather inefficient for large values of the box radius. The
other one leads to somewhat more complicated matrix elements but requires
smaller values of $N$ in order to obtain accurate results for large values
of $r_{0}$. In addition to producing quite accurate eigenvalues the RRM with
these basis sets proved useful for the study of the behaviour of the
spectrum of the model for very small and very large values of the box size.
We also obtained reasonably accurate eigenvalues by means of second-order
perturbation theory.

\begin{table}[tbp]
\caption{Eigenvalues for the even solutions when $r_{0}=1$, $\lambda =0.05$
and $i\leq N$}
\label{tab:e1_0.05}
\begin{center}
\par
\begin{tabular}{D{.}{.}{3}D{.}{.}{11}D{.}{.}{11}D{.}{.}{11}D{.}{.}{11}}
\hline \multicolumn{1}{l}{$N$}&\multicolumn{1}{c}{$E_1^{(e)}$}&
\multicolumn{1}{c}{$E_2^{(e)}$} &
\multicolumn{1}{c}{$E_3^{(e)}$} & \multicolumn{1}{c}{$E_4^{(e)}$} \\
\hline
  2 & 3.000033048 & 7.586264543 & 14.23335748 & 16.45868556   \\
  3 & 2.999983516 & 7.508025752 & 13.45385988 & 15.41429625   \\
  4 & 2.999940526 & 7.507311498 & 13.39694464 & 15.40025129   \\
  5 & 2.999940524 & 7.507179152 & 13.39164826 & 15.39159676   \\
  6 & 2.999940512 & 7.507178210 & 13.39154151 & 15.39154827   \\
  7 & 2.999940512 & 7.507178149 & 13.39153360 & 15.39153048   \\
  8 & 2.999940512 & 7.507178149 & 13.39153353 & 15.39153043   \\
  9 & 2.999940512 & 7.507178149 & 13.39153353 & 15.39153042   \\
 10 & 2.999940512 & 7.507178149 & 13.39153353 & 15.39153042   \\
\multicolumn{1}{l}{Ref. \cite{CAPF24}}  &2.999940  & 7.507196    & 13.391537   &              \\
 \end{tabular}
\par
\end{center}
\end{table}

\begin{table}[tbp]
\caption{Eigenvalues for the odd solutions when $r_0=1$, $\lambda=0.05$ and $%
i\leq N$}
\label{tab:o1_0.05}
\begin{center}
\begin{tabular}{D{.}{.}{3}D{.}{.}{11}D{.}{.}{11}D{.}{.}{11}D{.}{.}{11}}
\hline \multicolumn{1}{l}{$N$}&\multicolumn{1}{c}{$E_1^{(o)}$}&
\multicolumn{1}{c}{$E_2^{(o)}$} &
\multicolumn{1}{c}{$E_3^{(o)}$} & \multicolumn{1}{c}{$E_4^{(o)}$} \\
\hline
  3 & 7.507984721 & 13.45385988 & 22.77274842 & 25.97120953  \\
  4 & 7.507270472 & 13.39694464 & 20.61132495 & 24.8074732   \\
  5 & 7.507138138 & 13.39164826 & 20.59740341 & 24.78828653  \\
  6 & 7.507137196 & 13.39154151 & 20.58517127 & 24.77616193  \\
  7 & 7.507137135 & 13.3915336  & 20.58516517 & 24.77603555  \\
  8 & 7.507137134 & 13.39153353 & 20.58514439 & 24.77599977  \\
  9 & 7.507137134 & 13.39153353 & 20.58514439 & 24.77599951  \\
 10 & 7.507137134 & 13.39153353 & 20.58514438 & 24.77599947  \\
 11 & 7.507137134 & 13.39153353 & 20.58514438 & 24.77599947  \\
\multicolumn{1}{l}{Ref. \cite{CAPF24}}  & 7.507137    & 13.391537   &             &               \\

 \end{tabular}
\end{center}
\end{table}

\begin{table}[tbp]
\caption{First five eigenvalues for some values of $r_0$ and $\lambda$}
\label{tab:Er0lam}
\begin{center}
\par
{\tiny
\begin{tabular}{D{.}{.}{3}D{.}{.}{9}D{.}{.}{9}D{.}{.}{11}D{.}{.}{9}D{.}{.}{9}}

\multicolumn{6}{c}{$r_0=1$}\\

\multicolumn{1}{c}{$\lambda$} &                 &
&                  &                   &
\\
 0.05     &   2.999940513   &     7.507137135   &     7.507178149  &     13.39153353   &     13.39153353  \\
 0.50     &   2.994054784   &     7.503668344   &     7.507765819  &     13.39108541   &     13.39108591  \\
 1.00     &   2.976261687   &     7.493165697   &     7.509507984  &     13.38971472   &     13.38972271  \\
 1.50     &   2.946746472   &     7.475690676   &     7.512284352  &     13.38738763   &     13.38742819  \\
 2.00     &   2.905712640   &     7.451286862   &     7.515910615  &     13.38404044   &     13.38416912  \\
 2.50     &   2.853432944   &     7.420014281   &     7.520141153  &     13.37958431   &     13.37989994  \\
 3.00     &   2.790238712   &     7.381948462   &     7.524679955  &     13.37390580   &     13.37456372  \\
 3.50     &   2.716507923   &     7.337179290   &     7.529192808  &     13.36686757   &     13.36809331  \\
 4.00     &   2.632652970   &     7.285809695   &     7.533319836  &     13.35830959   &     13.36041264  \\
 4.50     &   2.539108909   &     7.227954221   &     7.536687533  &     13.34805070   &     13.35143815  \\
 5.00     &   2.436322782   &     7.163737522   &     7.538919695  &     13.33589085   &     13.34108027  \\
 5.50     &   2.324744395   &     7.093292825   &     7.539646842  &     13.32161397   &     13.32924491  \\
 6.00     &   2.204818732   &     7.016760401   &     7.538513927  &     13.30499156   &     13.31583500  \\
 \hline \multicolumn{6}{c}{$r_0 = 2$}\\
 0.05     &   1.121533076   &     2.471308473   &     2.471677189  &      4.092484156  &      4.092484253 \\
 0.50     &   1.055749927   &     2.425491724   &     2.461288707  &      4.079919270  &      4.080859951 \\
 1.00     &   0.8676905872  &     2.290990158   &     2.423192111  &      4.029818289  &      4.043205109 \\
 1.50     &   0.5825348562  &     2.079035409   &     2.345604043  &      3.921386314  &      3.974310688 \\
 2.00     &   0.2240418306  &     1.802394297   &     2.221750826  &      3.750830242  &      3.869531452 \\
 2.50     &  -0.1898793720  &     1.472757021   &     2.051207036  &      3.528077680  &      3.726784048 \\
 3.00     &  -0.6466661902  &     1.099674893   &     1.836953799  &      3.263614113  &      3.546435112 \\
 3.50     &  -1.137519525   &     0.6906132774  &     1.583000466  &      2.965401619  &      3.330401976 \\
 4.00     &  -1.656139155   &     0.2513481407  &     1.293278752  &      2.639409865  &      3.081315858 \\
 4.50     &  -2.197888537   &    -0.2136268090  &     0.9712902180 &      2.290243036  &      2.801986817 \\
 5.00     &  -2.759263211   &    -0.7007731576  &     0.6200711870 &      1.921520387  &      2.495126178 \\
 5.50     &  -3.337549806   &    -1.207265044   &     0.2422612717 &      1.536106044  &      2.163228074 \\
 6.00     &  -3.930602471   &    -1.730813281   &    -0.1598114054 &      1.136262144  &      1.808534720 \\
 \hline \multicolumn{6}{c}{$r_0 = 3$}\\
 0.05     &   1.000719415   &     2.013820524   &     2.013958910  &      3.057224946  &      3.057227051 \\
 0.50     &   0.8810998253  &     1.901568472   &     1.916966478  &      2.965602260  &      2.977378029 \\
 1.00     &   0.5302449413  &     1.575816276   &     1.649888731  &      2.680883025  &      2.751206873 \\
 1.50     &  -0.01534172516 &     1.074370141   &     1.261029544  &      2.240275829  &      2.407990915 \\
 2.00     &  -0.7078968358  &     0.4396646394  &     0.7809055363 &      1.681527101  &      1.968686336 \\
 2.50     &  -1.505217743   &    -0.2930063285  &     0.2246424266 &      1.035590763  &      1.446909507 \\
 3.00     &  -2.377395710   &    -1.098385329   &    -0.3990488847 &      0.3264125547 &      0.8529729025\\
 3.50     &  -3.304713209   &    -1.959183144   &    -1.084579031  &     -0.4273630821 &      0.1956863049\\
 4.00     &  -4.274142731   &    -2.863493879   &    -1.828291585  &     -1.211048117  &     -0.5170228781\\
 4.50     &  -5.276820708   &    -3.802913311   &    -2.626732853  &     -2.014028483  &     -1.277913746 \\
 5.00     &  -6.306499209   &    -4.771332759   &    -3.474706132  &     -2.830565891  &     -2.080445685 \\
 5.50     &  -7.358622358   &    -5.764184603   &    -4.365053586  &     -3.659193962  &     -2.918861635 \\
 6.00     &  -8.429766567   &    -6.777965138   &    -5.290258763  &     -4.500526982  &     -3.788232221 \\

 \end{tabular}
}
\par
\end{center}
\end{table}

\begin{table}[tbp]
\caption{$r_0^2E$ for $r_0=0.01$}
\label{tab:r0^2E}
\begin{center}
\par
\begin{tabular}{D{.}{.}{11}D{.}{.}{11}}
    \hline
\multicolumn{1}{c}{$r_0^2E$} & \multicolumn{1}{c}{$E^{PB}$}  \\
\hline
2.891592982  & 2.891592981 \\
7.340985322  & 7.340985321 \\
13.18730821  & 13.18730821 \\
15.23563117  & 15.23563117   \\
\end{tabular}
\par
\end{center}
\end{table}

\begin{table}[tbp]
\caption{Eigenvalue $E_{00}$ for increasing values of $r_0$ and some values
of $\lambda$}
\label{tab:E00(r0)}
\begin{center}
\par
\begin{tabular}{D{.}{.}{1}D{.}{.}{9}D{.}{.}{9}D{.}{.}{11}D{.}{.}{9}D{.}{.}{9}}

\hline \multicolumn{1}{c}{$r_0$} & \multicolumn{1}{c}{$\lambda=1$}
& \multicolumn{1}{c}{$\lambda=2$} &
\multicolumn{1}{c}{$\lambda=3$}& \multicolumn{1}{c}{$\lambda=4$}
&\multicolumn{1}{c}{$\lambda=5$}
\\ \hline
1      &  2.9762616872    &     2.9057126397    &    2.7902387123    &     2.6326529698    &  2.4363227822        \\
2      &  0.8676905872    &     0.2240418306    &   -0.6466661902    &    -1.6561391549    & -2.7592632114        \\
3      &  0.5302449413    &    -0.7078968358    &   -2.3773957102    &    -4.2741427314    & -6.3064992087        \\
4      &  0.5003869888    &    -0.9754437781    &   -3.2281562207    &    -5.9149227688    & -8.8273550313        \\
5      &  0.5000005484    &    -0.9996955349    &   -3.4775005103    &    -6.7378369178    & -10.4347543324       \\
6      &  0.5000000001    &    -0.9999995776    &   -3.4997274446    &    -6.97857932      & -11.24353629         \\
\multicolumn{1}{c}{$\infty$}  &0.5&    -1               &   -3.5
&    -7               &    -11.5

 \end{tabular}
\par
\end{center}
\end{table}

\begin{table}[tbp]
\caption{First two coefficients of the perturbation expansion for the lowest
energies when $r_0=1$}
\label{tab:PT_E_n}
\begin{center}
\par
\begin{tabular}{D{.}{.}{30}D{.}{.}{30}}
    \hline
\multicolumn{1}{c}{$E^{(0)}$} & \multicolumn{1}{c}{$E^{(2)}$}  \\
\hline
3.00000000000000000000000000000&  -0.0237951331606905590236056369694    \\
7.50717218045194296125597056932&  -0.0140183043464296058590713039161    \\
7.50717218045194296125597056932&   0.00238755957790725763825593365598   \\
13.3915380494648018553204940093&  -0.00180627777546582495721462311771   \\
13.3915380494648018553204940093&  -0.00180627777546582495721462311771   \\

\end{tabular}
\par
\end{center}
\end{table}

\begin{table}[tbp]
\caption{$E_{n,\nu}^{(0)}$ and $E_{n,\nu}^{(2)}$ for increasing values of $%
r_0$}
\label{tab:E^(2)_(nn)}\vspace{0.5cm}
\par
\begin{tabular}{rrrrrrr} \hline
\multicolumn{1}{c}{$r_0$} & \multicolumn{1}{c}{$E_{00}^{(e,0)}$} &
\multicolumn{1}{c}{$E_{10}^{(e,0)}$} & \multicolumn{1}{c}{$E_{20}^{(e,0)}$}
& \multicolumn{1}{c}{$E_{01}^{(e,0)}$} & \multicolumn{1}{c}{$E_{11}^{(e,0)}$}
& \multicolumn{1}{c}{$E_{02}^{(e,0)}$} \\ \hline
1 & 3.00000000 & 15.39153805 & 37.60583487 & 7.50717218 & 24.77601000 &
13.39153805 \\
2 & 1.12220853 & 4.44050521 & 10.01698219 & 2.47177521 & 6.82577445 &
4.09259935 \\
3 & 1.00193679 & 3.08886540 & 5.68860128 & 2.01496711 & 4.27163834 &
3.05805047 \\
4 & 1.00000336 & 3.00063504 & 5.02056599 & 2.00004978 & 4.00395689 &
3.00036610 \\
5 & 1.00000000 & 3.00000035 & 5.00003719 & 2.00000002 & 4.00000378 &
3.00000019 \\ \hline
\end{tabular}
\par
\begin{tabular}{rrrr}
\multicolumn{1}{c}{$r_0$} & \multicolumn{1}{c}{$E_{01}^{(o,0)}$} &
\multicolumn{1}{c}{$E_{11}^{(o,0)}$} & \multicolumn{1}{c}{$E_{02}^{(o,0)}$}
\\ \hline
1 & 7.50717218 & 24.77601000 & 13.39153805 \\
2 & 2.47177521 & 6.82577445 & 4.09259935 \\
3 & 2.01496711 & 4.27163834 & 3.05805047 \\
4 & 2.00004978 & 4.00395689 & 3.00036610 \\
5 & 2.00000002 & 4.00000378 & 3.00000019 \\ \hline
\end{tabular}
\par
\begin{tabular}{rrrrrrr}
\multicolumn{1}{c}{$r_0$} & \multicolumn{1}{c}{$E_{00}^{(e,2)}$} &
\multicolumn{1}{c}{$E_{10}^{(e,2)}$} & \multicolumn{1}{c}{$E_{20}^{(e,2)}$}
& \multicolumn{1}{c}{$E_{01}^{(e,2)}$} & \multicolumn{1}{c}{$E_{11}^{(e,2)}$}
& \multicolumn{1}{c}{$E_{02}^{(e,2)}$} \\ \hline
1 & -0.02379513 & -0.00304969 & -0.00116231 & 0.00238756 & 0.00085352 &
-0.00180628 \\
2 & -0.27022732 & -0.04303208 & -0.01639429 & -0.03917956 & 0.01445894 &
-0.04603071 \\
3 & -0.48698448 & -0.27520028 & -0.06592876 & -0.40340299 & -0.03096954 &
-0.32944152 \\
4 & -0.49995340 & -0.49391191 & -0.39406165 & -0.49914014 & -0.45745249 &
-0.49638216 \\
5 & -0.49999998 & -0.49999345 & -0.49945600 & -0.49999951 & -0.49990653 &
-0.49999640 \\ \hline
\end{tabular}
\par
\begin{tabular}{rrrr}
\multicolumn{1}{c}{$r_0$} & \multicolumn{1}{c}{$E_{01}^{(o,2)}$} &
\multicolumn{1}{c}{$E_{11}^{(o,2)}$} & \multicolumn{1}{c}{$E_{02}^{(o,2)}$}
\\ \hline
1 & -0.01401830 & -0.00420907 & -0.00180628 \\
2 & -0.18671149 & -0.06215340 & -0.04603071 \\
3 & -0.45868737 & -0.25480400 & -0.32944152 \\
4 & -0.49968212 & -0.48373136 & -0.49638216 \\
5 & -0.49999983 & -0.49996665 & -0.49999640 \\ \hline
\end{tabular}
\end{table}

\end{document}